\documentclass[twocolumn,aps,showpacs,prl,amsmath,amssymb,floatfix,superscriptaddress]{revtex4}
\usepackage{dcolumn}% Align table columns on decimal point
\usepackage{bm}% bold math
\usepackage{amsmath}
\usepackage{amsfonts}   
\usepackage{graphicx} 
 
\begin{document}
\title{Facilitated Asymmetric Exclusion}
\author{Alan Gabel}
\affiliation{Center for Polymer Studies and Department of Physics, Boston University, Boston,
Massachusetts 02215, USA}
\author{P. L. Krapivsky}
\affiliation{Center for Polymer Studies and Department of Physics, Boston University, Boston,
Massachusetts 02215, USA}
\author{S. Redner}
\affiliation{Center for Polymer Studies and Department of Physics, Boston University, Boston,
Massachusetts 02215, USA}

\begin{abstract}

  We introduce a class of facilitated asymmetric exclusion processes in which
  particles are pushed by neighbors from behind. For the simplest version in
  which a particle can hop to its vacant right neighbor only if its left
  neighbor is occupied, we determine the steady state current and the
  distribution of cluster sizes on a ring. We show that an initial density
  downstep develops into a rarefaction wave that can have a jump
  discontinuity at the leading edge, while an upstep results in a shock wave.
  This unexpected rarefaction wave discontinuity occurs generally for
  facilitated exclusion processes.

\end{abstract}

\maketitle

In the asymmetric exclusion process (ASEP), sites of a lattice are occupied
by single particles, each of which can hop at a fixed rate to a neighboring
vacant site on the right~\cite{L99,SZ95,D98,S00,BE07}.  This versatile model
describes many systems, including traffic~\cite{KS99,PS99,AS00,SCN10}, ionic
conductors~\cite{R77}, and RNA transcription~\cite{MGB68, MG69}. Despite its
simplicity, the properties of the ASEP are rich and deep.  For example, a
density that increases with $x$ leads to a propagating shock wave, similar to
a traffic jam that propagates along a congested road.  Conversely, when the
initial density drops quickly as a function of $x$, a rarefaction wave arises
in which the drop gradually smooths out, as occurs in stopped traffic after a
stoplight turns green.  Macroscopic aspects of these phenomena can be
understood from hydrodynamic theories~\cite{W74,KRB10}, while the
fluctuations about these macrostates continue to be actively
investigated~\cite{J00,PS02,TW09,BAC,CFP}.

\begin{figure}[ht]
\centerline{\includegraphics*[width=0.4\textwidth]{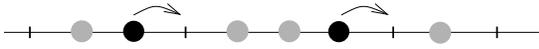}}
\caption{Illustration of occupancy facilitated asymmetric exclusion.
  Particles that are eligible to hop to the right are dark, while immobile
  particles are shaded.  This configuration contains islands of lengths 2, 3,
  and 1 (left to right).}
  \label{model}
\end{figure}

In this work, we investigate \emph{facilitated\/} asymmetric exclusion.  We
primarily focus on {\em occupancy facilitation} in which a particle can hop
to its vacant right neighbor only if its left neighbor is also occupied
(Fig.~\ref{model}).  This model was proposed by Basu and Mohanty~\cite{BM09}
in the context of non-equilibrium absorbing state phase transitions.  We also
investigate {\em distance facilitation} in which the rate at which a particle
hops to a vacant right site is a decreasing function of the distance between
a particle and its closest left neighbor.

The notion of facilitated exclusion is part of a general class of ASEP models
in which the hopping rate of a particle depends on more than just the
occupancy of the neighboring site~\cite{KS99,PS99,AS00,BM09,SZL03}.  For
example, in glassy dynamics the particle mobility decreases as the local
density increases~\cite{RS03}.  Conversely, the presence of nearby particles
may increase hopping rates; for example, in molecular motor models a moving
particle can exert a hydrodynamic force that pushes other particles
along~\cite{HPLEEE}.  Moreover, a subset of phase space in occupancy
facilitated exclusion can be mapped onto the ASEP of extended
objects~\cite{MGB68,MG69,SW98,AB99,SZL03,LC03}, a model that was formulated
to mimic the traffic of ribosomes along RNA.

In occupancy facilitation, a mean-field hypothesis for the current is
$J=\rho^2(1-\rho)$; the expression accounts for the presence of two particles
and one vacancy and represents a natural generalization of the current
$J=\rho(1-\rho)$ in the ASEP.  As we show below, the current in facilitated
exclusion actually has a very different density dependence.  We also develop
a hydrodynamic description for an initial density step and predict that a
rarefaction wave develops a discontinuity at the leading edge.  Finally, we
provide a general criterion to understand this unexpected phenomenon in the
framework of distance facilitation.\medskip

\noindent\emph{Finite Ring:}  We first determine the density dependence of 
the current on a finite ring in occupancy facilitation.  The key to
understanding the steady-state spatial distribution of particles is the
notion of \emph{islands}.  An island is a string of occupied sites that are
delimited at both ends by vacant sites (Fig.~\ref{model}).  Each hopping
event transforms a triplet $\bullet\bullet\circ$ into $\bullet\circ\bullet$.
Depending on the occupancy of the next site, the number of islands either
increases, $\bullet\bullet\circ\,\circ\to \bullet\circ\bullet\,\circ$, or
remains the same, $\bullet\bullet\circ\,\bullet\to
\bullet\circ\bullet\,\bullet$, but cannot decrease.  Thus the system
eventually reaches a state where the number of islands is maximal.

For $\rho\leq\frac{1}{2}$, the constraint that the number of islands can
never decrease ensures that the system eventually reaches a static state that
consists of immobile single-particle islands.  The approach to the final
state has a rich time dependence~\cite{GKR}, particularly in the marginal
case $\rho=\frac{1}{2}$ where the number of active particles asymptotically
decays as $t^{-1/2}$ (see also Refs.~\cite{RPV00,J05,S08}).

In the $\rho>\frac{1}{2}$ steady state, the requirement that the number of
islands is maximal ensures that adjacent vacancies must be separated by at
least one particle.  Furthermore, configurations that contain the maximal
number of islands are equiprobable.  Indeed, let $P(C)$ be the steady-state
probability of being in a maximal-island configuration $C$.  Then the
stationarity condition is
\begin{equation}
\label{stationary}
P(C)\sum_{C'}R(C\rightarrow C')=\sum_{C'}P(C')R(C'\rightarrow C)\,,
\end{equation}
where $R(C\rightarrow C')$ is the evolution rate from configuration $C$ to
$C'$. Since $R=1$ if an evolution step is allowed and $0$ otherwise, we need
to count the number of ways into and out of a configuration to solve
Eq.~\eqref{stationary}.

The evolution out of a configuration is triggered by triplets of the form
$\bullet\bullet\circ$ at the right edge of any island of length $\geq 2$. The
system can evolve into the configuration $C$ from another maximal-island
configuration by the process
$\bullet\bullet\circ\,\bullet\rightarrow\bullet\circ\bullet\,\bullet$.  This
evolution can only happen at the left edge of an island of length $\geq
2$. Hence there are an equal number of terms on both sides of
Eq.~\eqref{stationary}.  If $P(C)$ are equal for all configurations,
Eq.~\eqref{stationary} is clearly satisfied. Thus, all maximum island
configurations are equiprobable in the steady state.

\begin{figure}[ht]
\centerline{\includegraphics*[width=0.45\textwidth]{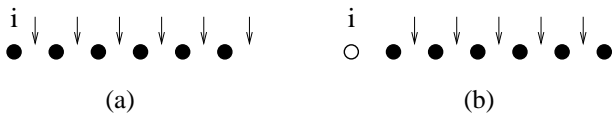}}
\caption{Illustration of number of places that $V$ vacancies can be placed
  among $N$ particles (filled circles) with: (a) site $i$ occupied or (b)
  vacant.}
  \label{choose}
\end{figure}

The probability of a maximum-island configuration therefore equals
$\mathcal{C}^{-1}$, where $\mathcal{C}$ is the total number of such
configurations with $N$ particles and $V$ vacancies on a ring of $L=N+V$
sites.  To determine $\mathcal{C}$, consider an arbitrary site that we label
by $i$.  If this site is occupied, there are $N$ possible locations between
the $N$ particles to put the $V$ vacancies (Fig.~\ref{choose}).  If site $i$
is unoccupied, there are $N-1$ possible places to put the remaining $V-1$
vacancies.  In both cases, we cannot put more than one vacancy between
consecutive particles or else the number of islands would not be maximal.
The number of such configurations is therefore given by
\begin{equation}
\label{w}
\mathcal{C} = \binom{N}{V} + \binom{N-1}{V-1}~.
\end{equation}

To obtain the steady state current, consider the flow across a link between
arbitrary adjacent sites $i$ and $i+1$.  For a particle to move across this
link, the consecutive sites $i-1$ and $i$ must be occupied while site $i+1$
must be vacant.  We now enumerate the number of maximum-island configurations
that are consistent with the presence of this triplet by noting that there
are $N-2$ places between the remaining particles to place the $V-1$ remaining
vacancies so that no two vacancies are adjacent.  Thus the number of allowed
configurations consistent with the presence of this triplet is
$\binom{N-2}{V-1}$.  The current across link $(i,i+1)$ is therefore
\begin{equation}
\label{j}
J = \frac{ \binom{N-2}{V-1}}{\mathcal{C}} \rightarrow \frac{(1-\rho)(2\rho-1)}{\rho}~,
\end{equation}
with $\rho=\frac{N}{L}$ held constant in the limit $N,L\rightarrow\infty$.
(This result can be mapped into an equivalent expression for the current in
the ASEP of extended objects~\cite{MGB68,MG69,SW98,AB99,SZL03,LC03}; we
return to this correspondence below.)  The current is zero at
$\rho=\frac{1}{2}$, since the system eventually reaches the static state of
alternating particles and vacancies.  The current is also zero at $\rho=1$
where no evolution is possible.  The maximal current arises when
$\rho^*=\frac{1}{\sqrt{2}}$, where $J(\rho^*)\equiv J_{\rm max}
=3-2\sqrt{2}\approx0.1716$.

\begin{figure}[ht]
\centerline{\includegraphics*[width=0.35\textwidth]{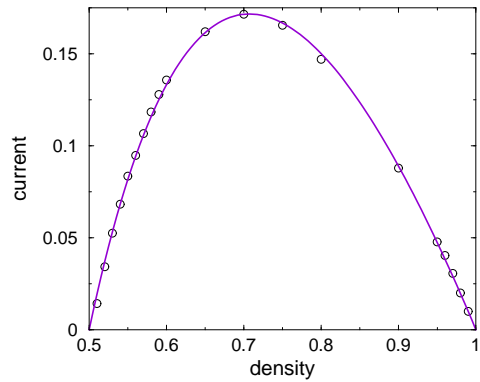}}
\caption{Current versus density for occupancy facilitation.  The smooth curve
  is the prediction \eqref{j}, while the circles are simulation data from
  $10^4$ realizations on a ring of $10^5$ sites.}
  \label{current}
\end{figure}

We can also determine $I_n$, the density of islands of length $n$.  Using the
same enumeration that gave the number of allowed configurations, there are
$V-2$ remaining vacancies that can be distributed among the $N-n-1$ places
between the rest of the particles so that there are no consecutive vacant
sites.  There are $\binom{N-n-1}{V-2}$ such configurations.  Since each
configuration has equal weight, the density of islands of length $n$ is
\begin{equation}
\label{In}
I_n=\frac{\binom{N-n-1}{V-2}}{\binom{N}{V}+\binom{N-1}{V-1}}\to
\frac{(1-\rho)^2}{\rho}\left(\frac{2\rho-1}{\rho}\right)^{n-1}~,
\end{equation}
where the latter equality applies for $n\ll L$; both Eqs.~\eqref{j} and
\eqref{In} were also derived in Refs.~\cite{SZL03,BM09} by independent
methods.  The island length distribution decays as $\lambda^n$, with
$\lambda=(2\rho-1)/\rho$, rather than $\lambda=\rho$ that occurs for a random
particle distribution.  Since $(2\rho-1)/\rho<\rho$, long islands are
suppressed compared to a random distribution; this feature is a consequence
of the constraint that the number of islands is maximal.  From these island
probabilities we recover the particle density from $\rho=\sum nI_n$, while
the current $J$ can alternatively be expressed as the probability to have an
island that contains at least two particles, $J=\sum_{n\geq2}I_n$.  \medskip

\noindent\emph{Density Step:}
Let us now study the evolution of a density step on the infinite line by
occupancy facilitation.  Initially, the density to the left of the origin is
$\rho_-$, while the density to the right is $\rho_+$.  For a downstep, where
$\rho_->\rho_+$, the density profile within a hydrodynamic description
evolves by the continuity equation $\frac{\partial \rho}{\partial
  t}+\frac{\partial J}{\partial x}=0$, which we may solve by the method of
characteristics~\cite{W74}.  The solution is a function of a scaled variable,
$z\equiv x/t$, so $\rho(x,t)=f(z)$.  Using the steady state current
expression \eqref{j} for the flux, we find that the scaled profile is
composed of distinct segments in which the density is either constant or
given by $f=(2+z)^{-1/2}$.  Thus the density profile is
\begin{equation}
\label{hydro}
f = 
\begin{cases}
\rho_-           &~\quad\quad\quad z<z_-\\
(2+z)^{-1/2} &\quad z_-<z<z_+\\
\rho_+          &~\quad\quad\quad z>z_+\,.
\end{cases}
\end{equation}
The position of the left interface $z_-$ is determined from continuity:
$(2+z_-)^{-1/2}=\rho_-$. When $\rho_->\rho^*=\frac{1}{\sqrt{2}}$, we have
$z_-<0$. In this situation, the density at the origin $\rho(0)$ is universal
and it coincides with the density $\rho^*$ that maximizes the current in
Eq.~\eqref{j}.  Therefore the number of particles that penetrates into the
region $x>0$ is $N(t)=J[\rho(0)]t = J_{\rm max} t$.

\begin{figure}[ht]
\centerline{\includegraphics*[width=0.375\textwidth]{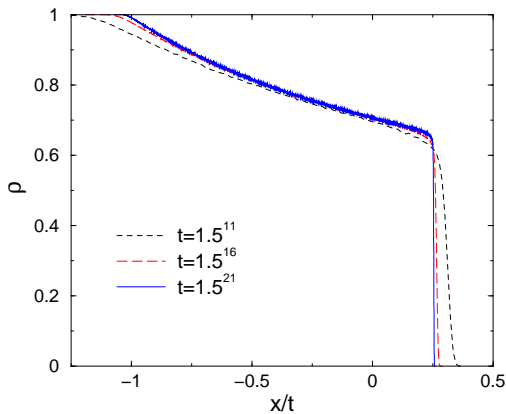}}
\caption{Scaled density profile of facilitated exclusion starting from the
  step initial condition $\rho_-=1$ and $\rho_+=0$.  The simulation data is
  based on $10^5$ realizations for three representative times and is visibly
  indistinguishable from the prediction of Eq.~\eqref{hydro} when
  $t=1.5^{21}$.}
  \label{rho}
\end{figure}

To locate the right interface $z_+$, we apply the constraint that the initial
mass within $[z_-,z_+]$ must equal the mass in this region at some later time
plus the net influx into this region.  In scaled units, this conservation
statement is
\begin{equation}
\label{cons}
  -\rho_-z_-+\rho_+ z_+ = \int_{z_-}^{z_+}\!\!\frac{dz}{\sqrt{2+z}}\,\,+ J_-- J_+\,,
\end{equation}
with $J_\pm = J(\rho_\pm)$. 
 
Different density profiles arise depending on whether $\rho_+<\frac{1}{2}$ or
$\rho_+>\frac{1}{2}$.  In the former case the right interface is located at
$z_+ = [2-3\rho_+ - 2\sqrt{(1-\rho_+)(1-2\rho_+)}]/\rho_+ ^2$.  As $z$ passes
through $z_+$ the density jumps from the value $(2+z_+)^{-1/2}$ to $\rho_+$.
For example, when $(\rho_-,\rho_+)=(1,0)$, $z_+=\frac{1}{4}$ and the
magnitude of the density drop is $\frac{2}{3}$ (Fig.~\ref{rho}).  The
discontinuity at the front of a rarefaction wave arises because the leading
particle cannot move unless ``pushed'' by neighboring particles from behind.
Consequently, the density at the leading edge must be non-zero.

For $\rho_+\geq\frac{1}{2}$, this jump discontinuity disappears, and the
density profile is everywhere continuous.  Continuity at $z=z_+$ now gives
$\rho_+=(2+z_+)^{-1/2}$, which manifestly solves Eq.~\eqref{cons}. For this
class of rarefaction waves, the density is sufficiently large ahead of the
wave that the leading edge can get pulled ahead and there is no need for a
pileup of particles from behind to push the wavefront forward.

To study shock waves, we suppose that $\frac{1}{2}<\rho_-<\rho_+$ and
consider a large region that includes the interface.  The particle influx to
this region is $J_-$, while the outflux is $J_+$.  The net flux must equal
the change in mass $c(\rho_--\rho_+)$ inside this region, where $c$ is the
shock wave speed.  Hence $c=(J_- - J_+)/(\rho_--\rho_+)$.  Using the
expression \eqref{j} for the current, the shock wave speed is
\begin{equation}
\label{c}
c= (\rho_-\rho_+)^{-1}-2\,.
\end{equation}
The shock propagates to the right if $\rho_-<(2\rho_+)^{-1}$ and to the left
otherwise.

Our results can be extended to a more stringent occupancy facilitation in
which $r$ consecutive sites to the left of a particle must be occupied for a
particle to hop to a vacant right neighbor~\cite{BM09,GKR}.  (For example,
for $r=3$ the rightmost particle in $\circ\bullet\bullet\bullet\circ$ cannot
move, while the update
$\bullet\bullet\bullet\bullet\circ\rightarrow\bullet\bullet\bullet\circ\bullet$
is possible.)~ A steady state with a non-vanishing current and a maximal
number of islands, each of length $\geq r$, arises when $\rho>\frac{r}{1+r}$.
All such configurations are again equiprobable.

\begin{figure}[ht]
\centerline{\includegraphics*[width=0.3\textwidth]{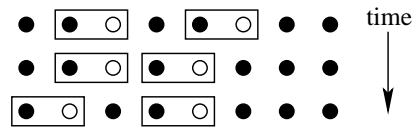}}
\caption{Equivalence between occupancy facilitated exclusion for the case
  $r=1$ and the ASEP of dimers (rectangles) that hop to the left.}
  \label{mapping}
\end{figure}

We now discuss the connection between occupancy facilitation and the ASEP of
extended objects \cite{MGB68,MG69,SW98,AB99,SZL03,LC03}.  In the
maximal-island steady-state regime with density $\rho\geq\frac{1}{2}$, we may
equivalently view a particle followed by a vacancy as an extended object of
length $k=2$ which hops to the \emph{left}.  Since vacancies cannot be
adjacent in the steady state, these extended objects obey exclusion and
perform a simple ASEP (Fig.~\ref{mapping}).  This connection continues to
hold with the more stringent $r$-tuple occupancy facilitation.  In the region of
phase space where $\rho\geq\frac{r}{r+1}$, the steady state behavior of the system 
maps to the ASEP of extended objects with length $k=1+r$.

Finally, we treat distance facilitation.  To find the current even for the
simple example in which the hopping rate equals $\ell^{-1}$, where $\ell$ is
the distance to the nearest left particle, is challenging.  The hydrodynamic
behavior, however, is robust and the rarefaction wave discontinuity always
arises~\cite{GKR}.  We can demonstrate the universality of this phenomenon
from basic features of the current-density relation.  We know that
$J(0)=J(1)=0$ and we expect that $J(\rho)$ has a single maximum at some
density $\rho^*$.  Applying the scaling ansatz for the continuity equation
shows that either $\rho$ is constant or $\frac{dJ}{d\rho}=z$.  The
rarefaction wave therefore has the form
\begin{equation*}
\rho(z)= 
\begin{cases}
\rho_-           & \,~\quad\qquad z<z_-\\
I(z) & \quad z_-<z<z_+\\
\rho_+          & \,~\quad\qquad z>z_+\,.
\end{cases}
\end{equation*}
where $I(z)$ is the inverse function of $z=\frac{dJ}{d\rho}$.
Differentiating this relation with respect to $z$ in the region $z_-<z<z_+$
gives $J_{\rho\rho}\rho_z=1$.  If $J_{\rho\rho}$ is everywhere negative (as
in the standard ASEP), then $\rho_z$ must also be negative.  Thus the density
$\rho(z)$ continuously decreases until it reaches $\rho_+$.
However, if $J_{\rho\rho}$ is positive at some low density, then $\rho_z$
would become positive.  Thus the smallest possible density $\rho_{\rm min}$
in a rarefaction wave occurs at the point where $J_{\rho\rho}$ vanishes.  If
$\rho_{\rm min}>\rho_+$, there must be a jump discontinuity at the leading
edge.

Thus an inflection point in the current-density relation signals a
rarefaction wave discontinuity.  Such an inflection point must exist for any
facilitation mechanism, since the $J_{\rho\rho}(\rho^*)<0$ at the maximum
$\rho^*$ and $J_{\rho\rho}>0$ for small $\rho$.  One such example is $J\sim
\rho^{\alpha+1}$ as $\rho\to 0$ that arises for distance facilitation with
hopping rate $\ell^{-\alpha}$.

In summary, facilitated asymmetric exclusion has features that are
dramatically different from simple asymmetric exclusion.  The most prominent
is the jump discontinuity at the leading edge of rarefaction waves.  This
phenomenon arises in a broad class of cooperative transport models with
facilitated dynamics.

\smallskip

We thank S. Grosskinsky, K. Jain, A. Schadschneider, and M. Sellitto for
helpful correspondence and literature advice.  We also gratefully acknowledge
financial support from NSF grant DMR-0906504 (AG and SR) and NSF grant
CCF-0829541 (PLK).


\begin{thebibliography}{99}

\bibitem{L99} T. M. Liggett, {\it Stochastic Interacting Systems: Contact,
    Voter, and Exclusion Processes} (Springer, New York, 1999).

\bibitem{SZ95} B. Schmittmann and R. K. P. Zia, 
%Statistical Mechanics of Driven Diffusive Systems, 
  in {\it Phase Transitions and Critical Phenomena}, Vol.\ 17, eds.\ C. Domb
  and J. L. Lebowitz (Academic Press, London, 1995).

\bibitem{D98} B. Derrida, Phys.\ Repts.\ {\bf 301}, 65 (1998);
 J. Stat.\ Mech.\ P07023 (2007).

\bibitem{S00} G. Sch\"utz, 
%Exactly Solvable Models for Many-Body Systems Far From Equilibrium, 
  in {\it Phase Transitions and Critical Phenomena}, Vol.\ 19, eds.\ C. Domb
  and J. L. Lebowitz (Academic Press, London, 2000).

\bibitem{BE07} R. A. Blythe and M. R. Evans, J. Phys.\ A {\bf 40}, R333
  (2007).

\bibitem{KS99} K. Klauck and A. Schadschneider, Physica A {\bf 271}, 102 (1999).

\bibitem{PS99} V. Popkov and G.M. Sch\"utz,  Europhys.\ Lett.\ {\bf 48}, 257
  (1999). 

\bibitem{AS00} T. Antal and G. M. Schutz, Phys.\ Rev.\ E {\bf 62}, 84 (2000).

\bibitem{SCN10} A. Schadschneider, D. Chowdhury, and K. Nishinari, {\it
    Stochastic Transport in Complex Systems: From Molecules to Vehicles},
  (Elsevier, 2010).

\bibitem{R77} P. M. Richards, Phys.\ Rev.\ B {\bf 16}, 1393 (1977).

\bibitem{MGB68} C. T. MacDonald, J. H. Gibbs, and A. C. Pipkin, Biopolymers {\bf 6}, 1
  (1968).

\bibitem{MG69} C. T. MacDonald and J. H. Gibbs, Biopolymers {\bf 7}, 707 (1969).

\bibitem{W74} G. B. Whitham, {\it Linear and Nonlinear Waves} (Wiley, New
  York, 1974).

\bibitem{KRB10} P. L. Krapivsky, S. Redner, and E. Ben-Naim, {\it A Kinetic
    View of Statistical Physics} (Cambridge University Press, Cambridge,
  2010).
  
\bibitem{J00} K. Johansson, Commun.\ Math.\ Phys.\ {\bf 209}, 437 (2000).

\bibitem{PS02} M. Pr\"ahofer and H. Spohn, in: {\it In and Out of Equilibrium}, 
ed.\ V. Sidoravicious (Birkh\"auser, Basel, 2002). 

\bibitem{TW09} C. A. Tracy and H. Widom, Commun.\ Math.\ Phys.\ {\bf 290},
  129 (2009); J. Math.\ Phys.\ {\bf 50}, 095204 (2009); J.\ Stat.\ Phys.\ {\bf
    137}, 825 (2009).

\bibitem{BAC}  G. Ben Arous and I. Corwin, arXiv:0905.2993.

\bibitem{CFP}  I. Corwin, P. L. Ferrari, and S. P\'ech\'e, J.\ Stat.\ Phys.\ {\bf 140}, 
232 (2010).

\bibitem{BM09} U. Basu and P. K. Mohanty,  Phys.\ Rev.\ E {\bf 79}, 041143 (2009).

\bibitem{SZL03} L. B. Shaw, R. K. P. Zia, and K. H. Lee, Phys.\ Rev.\ E {\bf
    68}, 021910 (2003).

\bibitem{RS03} A review of this topic is given in F. Ritort and P. Sollich,
  Adv.\ Phys.\ {\bf 52}, 219 (2003).

\bibitem{HPLEEE} D. Houtman et al., Europhys.\ Lett.\ {\bf 78}, 18001 (2007).

\bibitem{AB99} F. C. Alcaraz and R. Z. Bariev, Phys.\ Rev.\ E {\bf 60}, 79 (1999).

\bibitem{SW98} T. Sasamoto and M. Wadati, J. Phys.\ A {\bf 31}, 6057 (1998).

\bibitem{LC03} G. Lakatos and T. Chou, J. Phys.\ A {\bf 36}, 2027 (2003).

\bibitem{GKR} A. Gabel, P. L. Krapivsky, and S. Redner, in preparation.

\bibitem{RPV00} M.~Rossi, R.~Pastor-Satorras, and A.~Vespignani, Phys.\ Rev.\
  Lett.\ {\bf 85}, 1803 (2000).
     
\bibitem{J05} K. Jain, Phys.\ Rev.\ E {\bf 72}, 017105 (2005).

\bibitem{S08} M.~Sellitto, Phys.\ Rev.\ Lett.\ {\bf 101}, 048301 (2008).

\end{thebibliography}
\end{document}